\documentclass[amsmath,amsfonts,twocolumn,prl]{revtex4-1}
\usepackage{graphicx}
\usepackage{epsfig}
\usepackage{bm}
\usepackage{dcolumn}
\usepackage{amsmath}
\usepackage{amssymb}
\usepackage{color}

\begin{document}

\title{Temperature Dependence of the Anomalous Hall Effect from Electron Interactions}

\author{Songci Li}
\affiliation{Department of Physics, University of Wisconsin--Madison, Madison, Wisconsin 53706, USA}

\author{Alex Levchenko}
\affiliation{Department of Physics, University of Wisconsin--Madison, Madison, Wisconsin 53706, USA}

\begin{abstract}
We consider the impact of electron-electron interactions on the temperature dependence of the anomalous Hall effect in disordered conductors. The microscopic analysis is carried out within the diagrammatic approach of the linear response Kubo-Streda formula with an account of both extrinsic skew-scattering and side-jump mechanisms of the anomalous Hall effect arising in the presence of spin-orbit coupling. We demonstrate the importance of electron interactions in the Cooper channel even for nominally non-superconducting materials and find that the corresponding low-temperature dependence of the anomalous Hall conductivity is asymptotically of the form $\sqrt{T}/\ln(T_0/T)$ in three dimensions and $\ln[\ln(T_0/T)]$ in two dimensions, where the scale of $T_0$ is parametrically of the order of Fermi energy. These results, in particular, may provide a possible explanation for the recently observed unconventional temperature dependence of the anomalous Hall effect in HgCr$_2$Se$_4$.    
\end{abstract}

\date{February 28, 2020}

\maketitle

\textit{Introduction}.--The anomalous Hall effect (AHE) is incredibly rich and complex transport phenomenon, see Refs.~\cite{Sinova-RMP,Sinitsyn-JPCM,Niu-RMP} for comprehensive reviews and references herein. The key ingredients of AHE are the broken time-reversal and spin-rotational symmetries, and one usually distinguishes between intrinsic and extrinsic mechanisms of anomalous conduction. The intrinsic mechanism, termed as anomalous velocity, was discovered by Karplus and Luttinger~\cite{KL}, and arises from the transverse drift of electrons moving in a perfect periodic lattice subject to spin-orbit coupling. Importantly, it can be interpreted in terms of the Berry phase associated with the motion of Bloch electrons in momentum space~\cite{Chang-Niu}. The topological origin of the anomalous velocity can lead to a quantized anomalous Hall effect and the  corresponding conductance can be expressed in terms of the integral of the Berry curvature over the momentum space or the Chern number for fully filled bands. In a generic disordered system, the quantum nature of electron scattering by impurities combined with strong spin-orbit effects leads to a right-left asymmetry in the differential scattering cross-section of the average scattering probability. This extrinsic effect was found by Smit~\cite{Smit} and is known as the skew-scattering mechanism. Technically it appears to the third order in the scattering potential, namely beyond the leading Born approximation. In addition, impurity scattering also leads to a coordinate shift in electron trajectories that gives rise to an extra contribution to the velocity operator. This mechanism was revealed by Berger~\cite{Berger} and is termed a side-jump accumulation. In the semiclassical approaches to AHE based on the Boltzmann equation, one also discusses the so-called anomalous distribution mechanism, however this term is in fact just a part of the side-jump process in the language of Kubo-Streda formulas. The connection between diagrammatic and kinetic equation approaches has been discussed in Ref.~\cite{Sinitsyn-BKE-KS}, including the recent discussion of remaining discrepancies between them in relation to diffractive skew-scattering~\cite{Ado-1,Konig-AHE-KTI,Konig-Kerr,Ado-2}. Unlike the anomalous velocity term, extrinsic mechanisms are obviously not universal as, in particular, they depend on the statistical properties of disorder and strength of the scattering potential itself. It is of interest to point out, however, that in the case of centro-symmetric impurity potential electron coordinate shifts upon scattering can be expressed solely in terms of Bloch functions associated with the motion of electrons in a periodic potential of the crystal. The latter can in turn be related to the Pancharatnam phase which represents a special case of Berry phase~\cite{Belinicher,Sinitsyn-SJ}. In this limit side-jump mechanism becomes universal and can be regarded, in some sense, as intrinsic. The concept of side-jump remains even in the pure system and can be associated with the two-particle collisions processes. This gives rise to the hydrodynamic limit of AHE~\cite{Pesin}. 

Experimentally, extrinsic mechanisms can be distinguished by their respective scalings with impurity concentration, and thanks to recent advances in control of impurity density, comprehensive scaling between the anomalous Hall conductivity and longitudinal conductivity has been established, see for example Refs.~\cite{Jin-PRL09,Zhao-PRB14,Jin-PRL15} for the detailed discussion. However, substantial complications arise in analysis of the data since low-temperature transport properties of disordered systems are strongly affected by quantum corrections. These are most prominently weak-localization (WL) and electron-electron interaction (EEI) terms. In the context of longitudinal conductivity and conventional Hall effect, both of these quantum effects have been meticulously studied and well understood with, in general, excellent agreement between measurements and theoretical predictions~\cite{Bergmann,AA,Lee}. However, much less is known concerning the impact of quantum interference corrections to anomalous Hall transport, and especially in regards to the role of EEI effects.   

\textit{Overview and motivation}.--In the first detailed measurements carried out by Bergmann and Ye~\cite{Bergmann-Ye}, anomalous Hall conductivity of amorphous ferromagnetic thin films of Fe showed no noticeable temperature dependence in the quantum regime. In contrast, the corresponding anomalous Hall resistivity followed logarithmic temperature dependence similar to Coulomb anomaly observed in the longitudinal resistance but twice its magnitude.  Since quantum coherent processes responsible for WL corrections were expected to be strongly suppressed by magnetic scatterings, and the only low-$T$ anomaly could be due to Coulomb interaction, the data was interpreted as EEI do not affect anomalous Hall conductivity. This conclusion was further supported by the theoretical analysis of Langenfeld and W\"olfle~\cite{Langenfeld} who showed by an explicit calculation that Coulomb anomaly terms due to Al'tshuler-Aronov (AA) corrections cancel identically from the anomalous Hall conductivity. The subsequent measurements on amorphous Fe$_x$Si$_{1-x}$ multilayers revealed more complex picture where not only anomalous contribution to Hall resistance was seen, but also a clear temperature dependence of the anomalous Hall conductance was observed~\cite{Valles}. The multitude of experiments that followed on a variety of material systems including (amorphous/polycrystalline/granular) ferromagnetic thin films of Fe, Ni, FePt, CoFeB, CNi$_3$~\cite{Mitra-Fe,Adams,Zhang-FePt,Jin-Fe,Zhang-Ni} as well as ferromagnetic semiconductors Ga$_{1-x}$Mn$_x$As and HgCr$_2$Se$_4$~\cite{Mitra-GaMnAs,Li-HgCrSe} provided more comprehensive evidences for the quantum anomalies in the temperature and disorder dependence of the anomalous Hall conductivity. These measurements in part also triggered multiple theoretical studies where the effects of WL and Coulomb interaction were re-analyzed~\cite{Dugaev,Wolfle,Muttalib,Kharitonov}. It was shown that in contrast to the conventional Hall effect, there exists a nonvanishing WL correction to the anomalous Hall resistivity, $\delta\rho_{xy}/\rho_H=[\delta\sigma_{xy}/\sigma_{xy}-2\delta\sigma_{xx}/\sigma_{xx}]$, where $\rho_H$ is the classical Hall resistance. While $\delta\sigma^{\mathrm{WL}}_{xy}$ vanishes in the side-jump mechanisms, it is finite for the skew-scattering. As a result, the total WL correction to $\sigma_{xy}$ does not cancel with the corresponding WL correction to $\sigma_{xx}$. Coulomb anomaly from direct and exchange terms were shown to be zero for the anomalous Hall conductivity \cite{Muttalib}, so that on top of WL, the additional $T$-dependence of anomalous Hall resistivity comes from the AA corrections to the diagonal conductivity, $\delta\sigma^{\mathrm{AA}}_{xx}$. Since in two-dimensions both $\delta\sigma^{\mathrm{WL}}$ and $\delta\sigma^{\mathrm{AA}}$ are logarithmic in temperature, it is a challenge to separate their relative importance in $\delta\rho_{xy}(T)$. 

\begin{figure}
\includegraphics[width=0.4\textwidth]{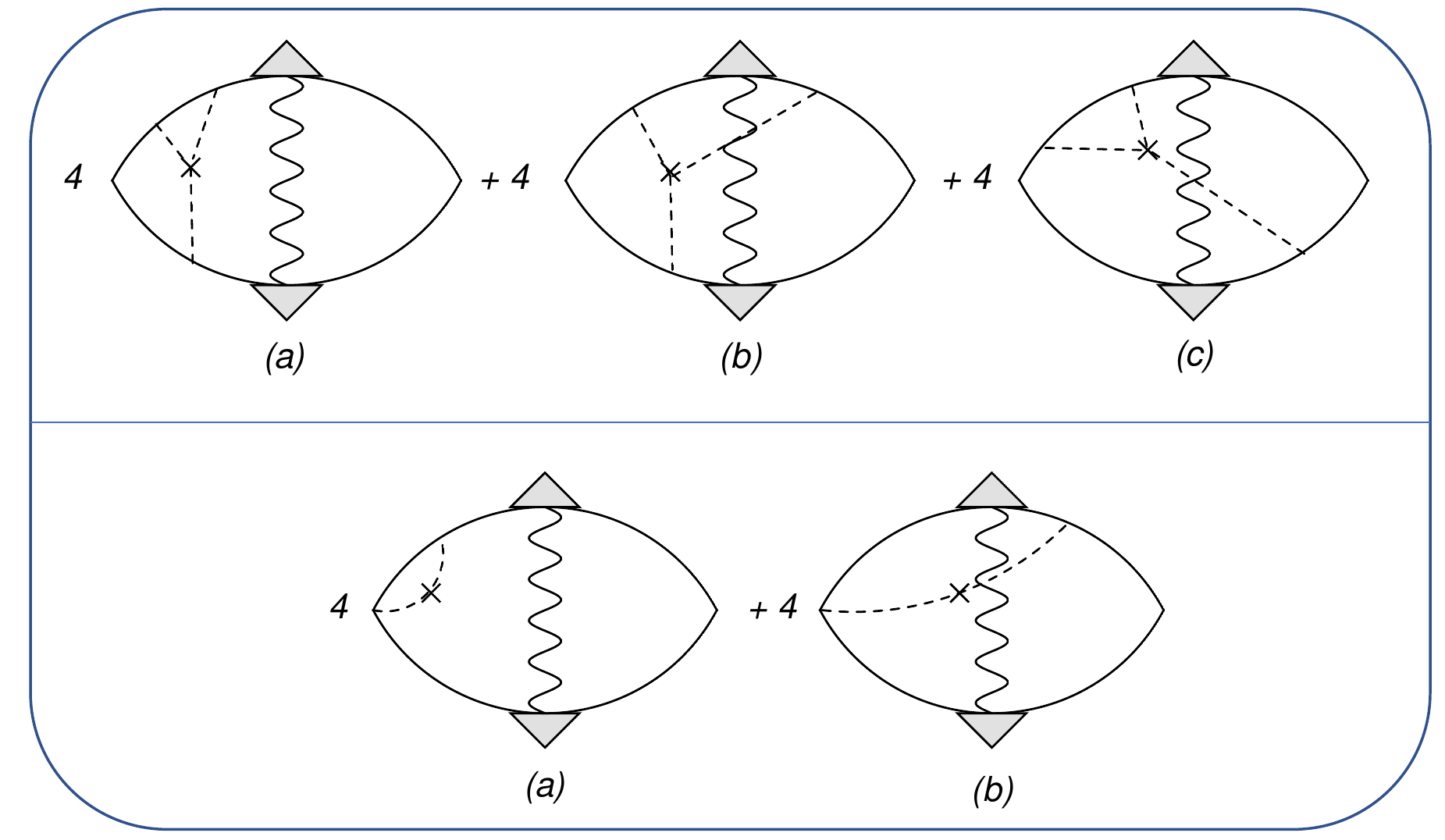} 
\caption{Diagrams for Maki-Thompson corrections in the skew scattering (upper panel) and side-jump (lower panel) mechanisms. The wavy line represents the EEI propagator in the Cooper channel Eq. \eqref{L}. The shaded triangle is the three-leg vertex (cooperon) Eq. \eqref{lambda}. Each diagram comes with four different copies corresponding to all possible arrangements of impurity lines for the amplitude of Eq. \eqref{V}. The single impurity line in the side-jump diagrams contains the accumulation velocity term of Eq. \eqref{v-sj} in the current vertex.} 
\label{Fig-MT}
\end{figure} 

The seemingly emergent conclusion from all the existing theories that the sole mechanism of $T$-dependence of $\delta\sigma^{\mathrm{AHE}}_{xy}$ from quantum interference processes is due to WL is at odds with various experimental facts and the most recent experimental results of AHE measurements in HgCr$_2$Se$_4$~\cite{Li-HgCrSe} in particular. This material is in the three-dimensional limit where WL and AA corrections have distinct temperature dependence. Indeed, provided that the leading source of decoherence is governed by the Coulomb interaction, which implies the dephasing time of the form $\tau^{-1}_{\phi}\simeq T\sqrt{T\tau}/(p_Fl)^2$, the WL correction $\delta\sigma^{\mathrm{WL}}\simeq\sigma_Q/\sqrt{D\tau_\phi}$ scales as $T^{3/4}$, where $\sigma_Q=e^2/\hbar$ is quantum of conductance, $\tau$ and $l=v_F\tau$ are the disorder mean free time and path, respectively, $D=v_Fl/3$ is the diffusion coefficient, and $p_F$ is the Fermi momentum. In contrast, the AA correction is of the form $\delta\sigma^{\mathrm{AA}}\simeq \sigma_Q(1-9F_\sigma/8)\sqrt{T/D}$,  where $F_\sigma$ is the triplet channel interaction constant (in this formula we took $F_\sigma\ll1$ for brevity). Based on these results the expectation is then that $\delta\rho^{\mathrm{AHE}}_{xy}\propto\sqrt{T}$ and $\delta\sigma^{\mathrm{AHE}}_{xy}\propto T^{3/4}$ but this is not what was seen experimentally. Measurements showed that the anomalous Hall conductivity also scales as $\sqrt{T}$~\cite{Li-HgCrSe}. This motivates the current work aimed at resolving the discrepancy between theory and experiment in regards to the effect of electron-electron interaction on the AHE. 
                
The key insight that one can try to explore, which was overlooked in the previous studies of AHE, comes from the work of Larkin~\cite{Larkin} who argued that low-temperature transport characteristics of disordered conductors can be strongly affected by the Cooper channel even for the case of purely repulsive interaction, namely for materials that do not undergo superconducting transition. As a guiding example Larkin demonstrated how WL corrections are modified with an account of Maki-Thompson contributions known from the context of superconducting fluctuations~\cite{Maki,Thompson}. This idea was further systematically developed by Al'tshuler \textit{et al.}~\cite{AVR} in a detailed study where all ten classes of leading ladder diagrams were carefully examined and temperature-depended corrections to diagonal conductivity were derived. It is the intent of this work to extend fundamentals of the theory of quantum interaction corrections~\cite{Larkin,AVR} to the case of the anomalous Hall transport phenomena.   

\begin{figure}
\includegraphics[width=0.45\textwidth]{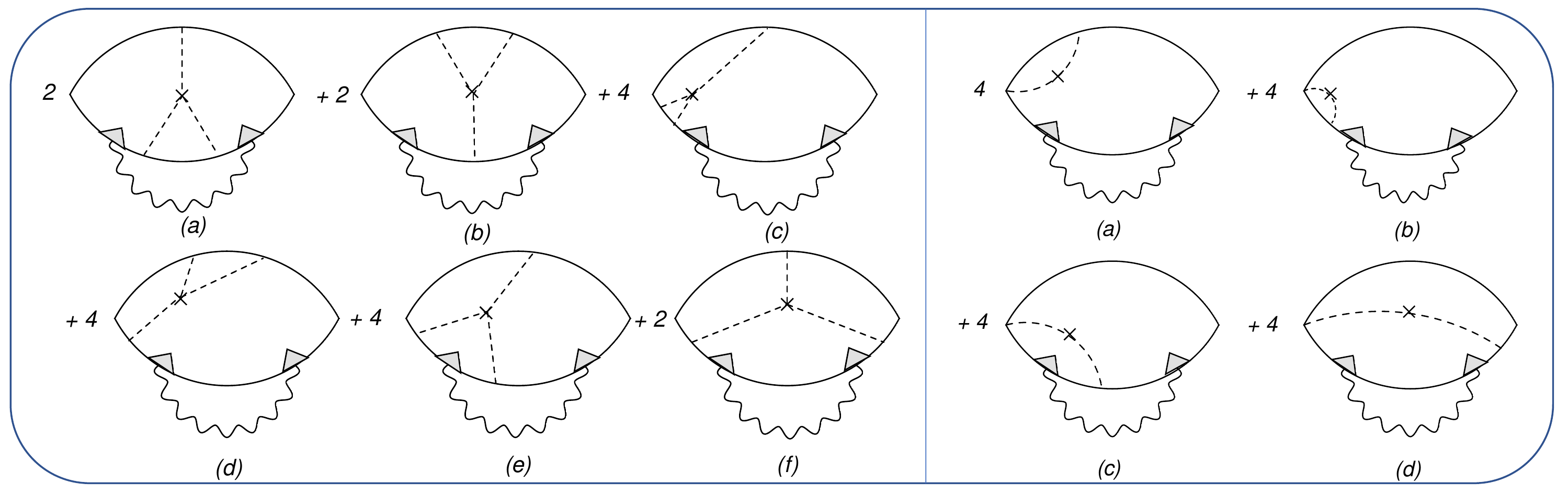}
\caption{Diagrams for the DOS corrections to the AHE in the skew scattering (left panel) and side-jump (right panel) mechanisms. The nomenclature is the same as in Fig. \ref{Fig-MT}.} 
\label{Fig-DOS}
\end{figure} 

\textit{Model}.--We adopt the same model as in Ref.~\cite{Langenfeld}, i. e. a model of spin unpolarized electrons scattering off localized magnetic moments with spin-orbit coupling. We perform calculations of AHE diagrammatically based on the linear-response Kubo-Streda formula. The elements in the diagrams are given as follows, see Figs.~\ref{Fig-MT} and~\ref{Fig-DOS}. The solid line denotes the impurity averaged Green's function of electrons
\begin{equation}\label{G}
G_{\bm{p}}(\varepsilon_n)=[i\widetilde{\varepsilon}_n-\xi_{\bm{p}}]^{-1}, \quad \widetilde{\varepsilon}_n=\varepsilon_n+\frac{1}{2\tau}\text{sgn}({\varepsilon_n}),
\end{equation}
where $\xi_{\bm{p}}=p^2/2m-E_F$ and $\varepsilon_n=(2n+1)\pi T$ is the Matsubara frequency. In the case of a disordered metal, $T \ll \tau^{-1} \ll E_F$, the three-leg cooperon vertex representing the sum of impurity ladder diagrams, is written in the form
\begin{equation}\label{lambda}
	\lambda_{\bm{q}}(\varepsilon_n,\varepsilon^\prime_n)=\frac{|\widetilde{\varepsilon}_n-\widetilde{\varepsilon}^\prime_n|}{|\varepsilon_n-\varepsilon^\prime_n|+Dq^2\Theta(-\varepsilon_n\varepsilon^\prime_n)},
\end{equation}
where $\Theta(x)$ is the Heaviside step-function. For the case of repulsive interactions in the Cooper channel, positive coupling constant $g>0$, the effect of EEI is captured by the propagator \cite{Larkin,AVR}
\begin{equation}\label{L}
L^{-1}_{\bm{q}}(\Omega_k)=\nu_F\Big[\ln\frac{T_0}{T}-\psi\left(\frac{1}{2}+\frac{|\Omega_k|}{4\pi T}+\frac{Dq^2}{4\pi T}\right)+\psi\left(\frac{1}{2}\right)\Big],
\end{equation}
which is depicted by the wavy line in diagrams. Here $\nu_F$ is the density of state on the Fermi surface, $T_0=E_F\exp(1/\nu_F g)$, and $\psi(x)$ is the digamma function. Lastly, the dashed line represents the impurity potential defined by the amplitude of scattering from the state with momentum $\bm{p}$ to $\bm{p}'$
\begin{equation}\label{V}
V_{\bm{pp}'}=V_0\left[1-\frac{i\alpha_{\text{so}}}{p^2_F}(\bm{p}\times\bm{p}')_z\right],
\end{equation}
where $\alpha_{\text{so}}$ is the dimensionless spin-orbit coupling constant giving rise to skew scattering and side jump processes. In the dc-limit, $\omega\to0$, conductivity can be found from the retarded component of the electromagnetic response function, $\sigma=Q^R(\omega)/(-i\omega)$. 

\begin{figure*}[t!]
\includegraphics[width=0.3\linewidth]{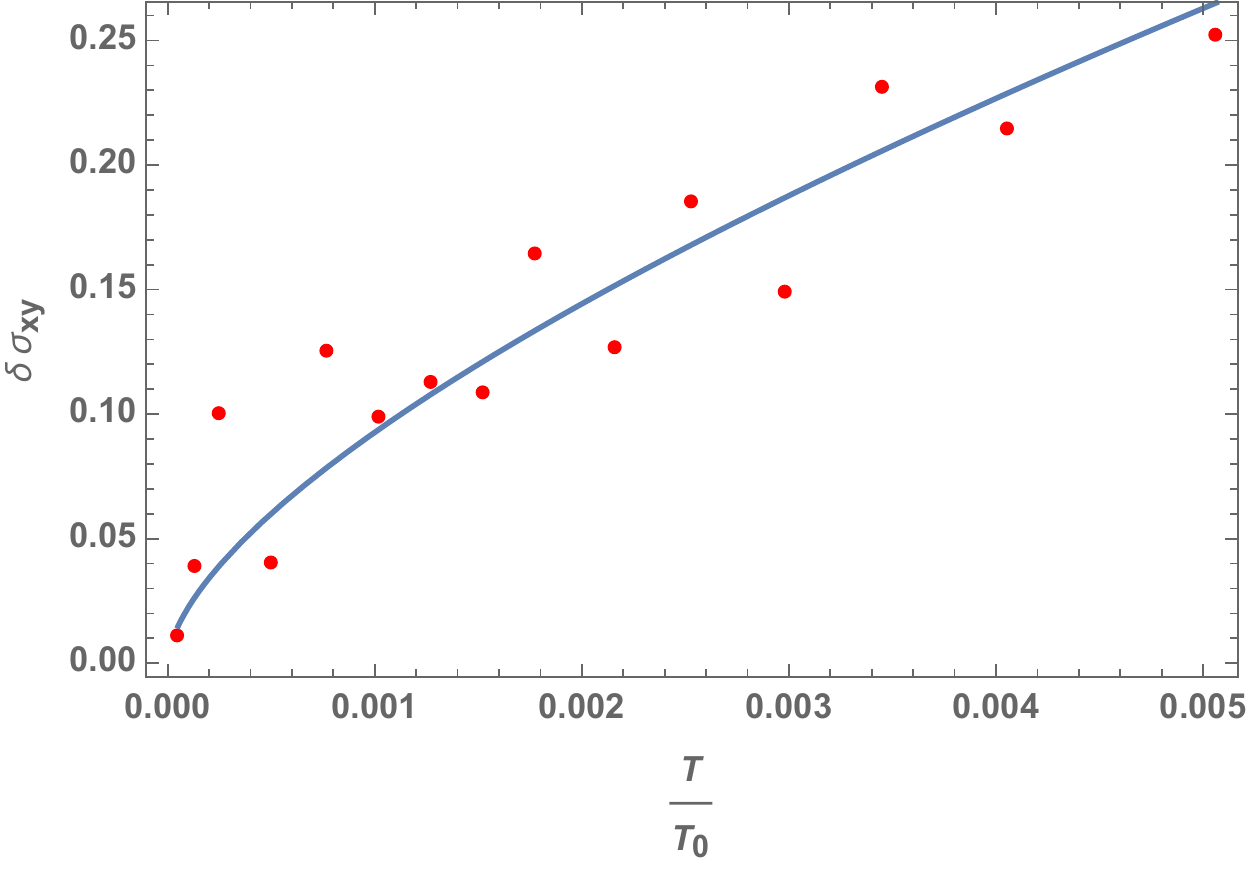}
\includegraphics[width=0.3\linewidth]{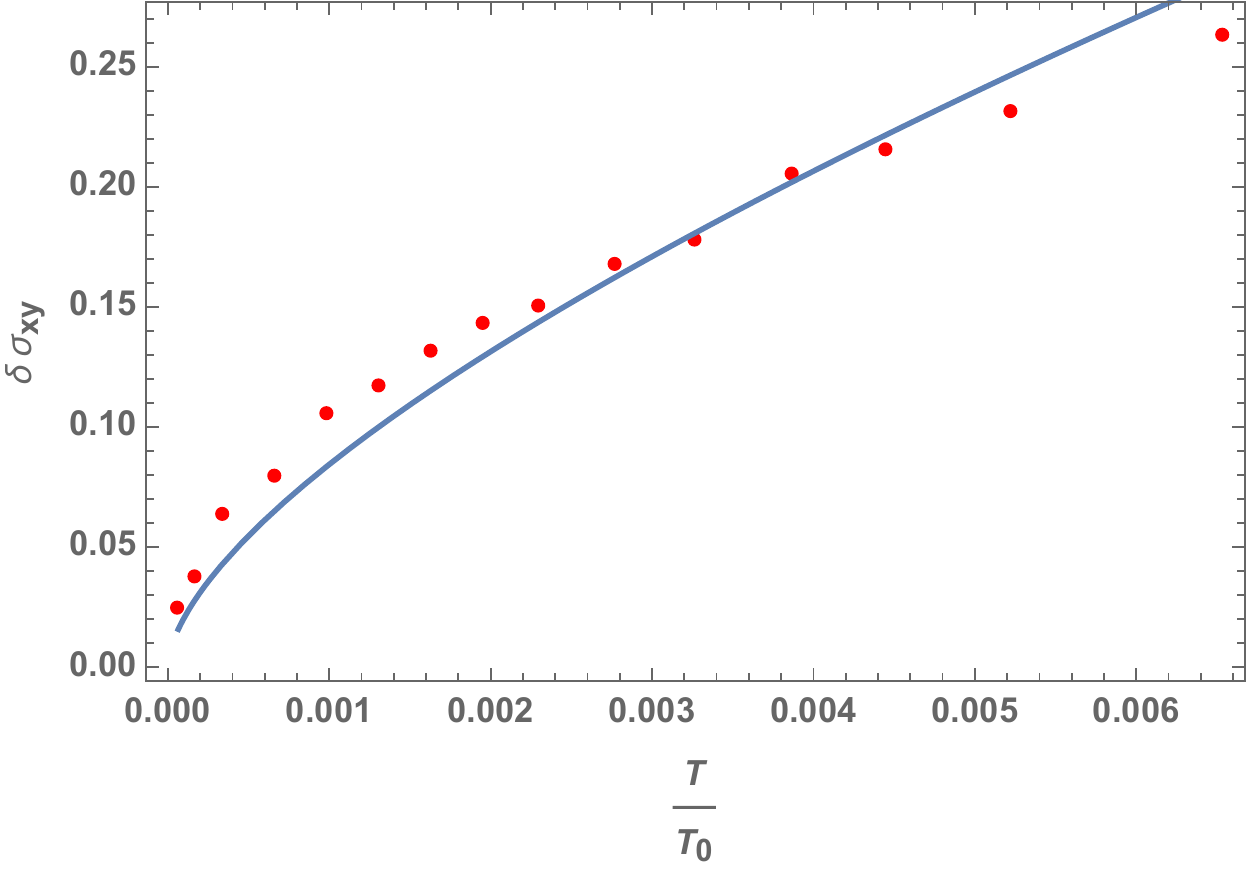}
\includegraphics[width=0.3\linewidth]{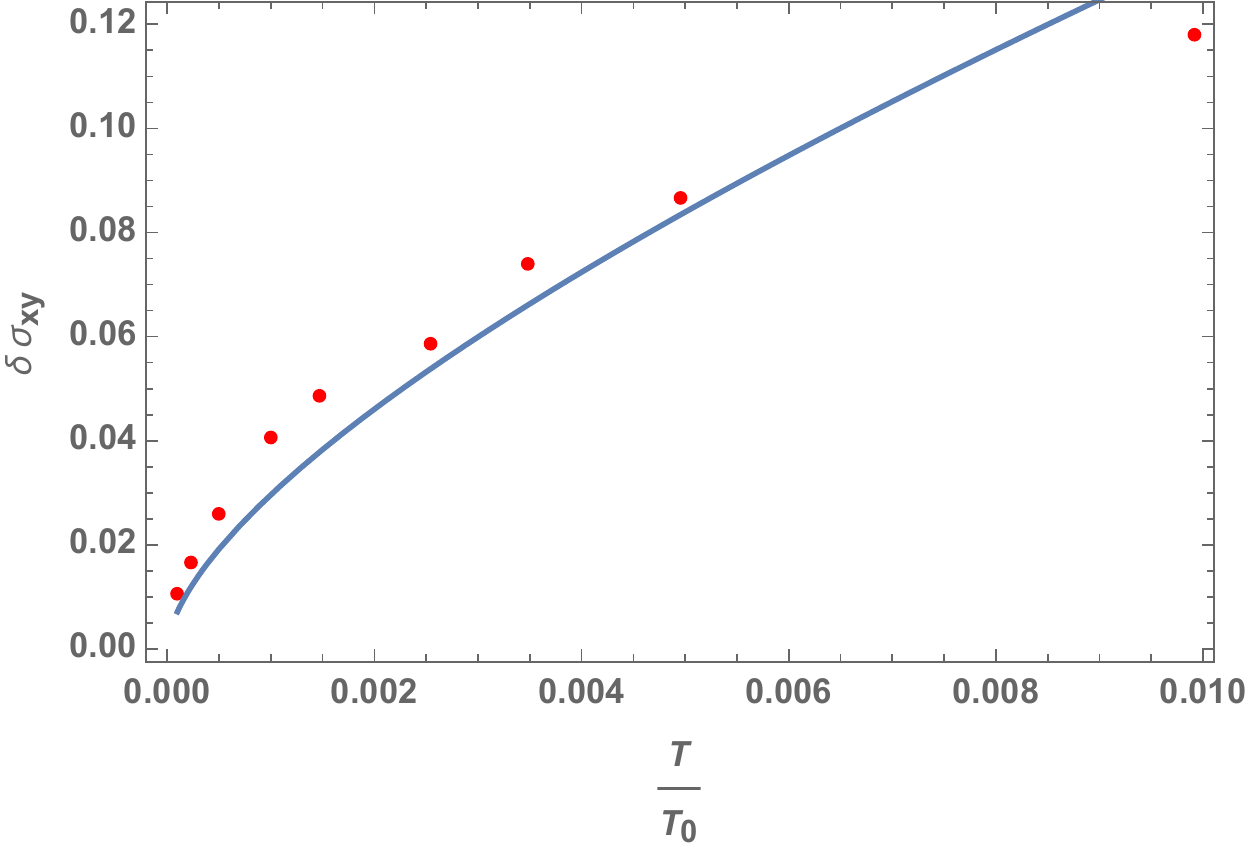}
\caption{Temperature dependence of the anomalous Hall conductivity for three different samples per measurements of Ref. \cite{Li-HgCrSe}, see their Fig. 3(d) and supplementary materials for further details. Solid line represents a theoretical fit to the calculated function $\delta\sigma_{xy}=\sigma_0\sqrt{t}[\ln^{-1}(1/t)-0.78\ln^{-2}(1/t)]$, where $t=T/T_0$, derived from Eqs. \eqref{sigma-sk-MT} and \eqref{sigma-sk-DOS}. The prefactor $\sigma_0\simeq 2.14\sigma_Q(\tau/\tau_{\mathrm{sk}})\sqrt{T_0/D}$ was used as a fitting parameter while $T_0$ was estimated for the given sample carrier concentration from Table I of Ref. \cite{Li-HgCrSe}. The scale of $\delta\sigma_{xy}$ is in units (Ohm cm)$^{-1}$.}  
\label{Fig-sigma}	
\end{figure*}

\textit{Skew-scattering mechanism}.--To the first order in interaction, and with account for impurity averaging, $Q(\omega)$ is given by ten distinct diagrams representing Aslamazov-Larkin (AL), Maki-Thompson (MT), and density of states (DOS) terms \cite{Book}. For the purpose of AHE calculation, each diagram needs to be generalized to include skew-scattering and side-jump effects. We begin with the skew-scattering mechanism and consider first MT diagrams, see the upper panel of Fig.~\ref{Fig-MT}. The analytical structure of the response kernel for all these diagrams has a common form   
\begin{equation}\label{Q}
Q_{xy}(\omega_\nu)=2e^2T\sum_{\Omega_k}\int\frac{d^dq}{(2\pi)^d}L_{\bm{q}}(\Omega_k)\Sigma_{\bm{q}}(\Omega_k,\omega_\nu)
\end{equation}
with the different expressions for $\Sigma_{\bm{q}}$. For example, the diagram-(a) reads explicitly 
\begin{equation}
\Sigma^{\text{sk-MT-a}}_{\bm{q}}=T\sum_{\varepsilon_n}\lambda_{\bm{q}}(\varepsilon_n,\Omega_{k-n})
\lambda_{\bm{q}}(\varepsilon_{n+\nu},\Omega_{k-n-\nu})J^{\text{sk-MT-a}}_{xy},
\end{equation} 
where $\varepsilon_{n+\nu}\equiv\varepsilon_n+\omega_\nu, \Omega_{k-n}\equiv\Omega_k-\varepsilon_n$, and we assume $\omega_\nu>0$ without loss of generality. The current block
\begin{align}
&J^{\text{sk-MT-a}}_{xy}=n_{\text{imp}}\nu^3_F\langle v_{\bm{p},x} v_{-\bm{k},y}V_{\bm{p}\bm{k}'}V_{\bm{k}'\bm{k}}V_{\bm{k}\bm{p}} \rangle\nonumber \\
& \times\int d\xi_{\bm{k}} G_{\bm{k}}(\varepsilon_n)G_{\bm{k}}(\varepsilon_{n+\nu})G_{-\bm{k}}(\Omega_{k-n-\nu})G_{-\bm{k}}(\Omega_{k-n}) \nonumber \\
& \times\int d\xi_{\bm{p}} d\xi_{\bm{k}'} G_{\bm{p}}(\varepsilon_n)G_{\bm{p}}(\varepsilon_{n+\nu})G_{\bm{k}'}(\varepsilon_{n+\nu})
\end{align}
contains an angular average $\langle\ldots\rangle$ over the directions of momenta, and $n_{\text{imp}}$ is the impurity concentration. Carrying out integrals over fermionic dispersions, angular average on the Fermi surface, and frequency summations followed by an analytical continuation, $\omega_\nu\to-i\omega$, one finds the  corresponding conductivity 
\begin{equation}
\sigma^{\text{sk-MT-a}}_{xy} = -\frac{e^2}{18\sqrt{\pi}}\frac{\tau}{\tau_{\mathrm{sk}}}\int\limits^{\infty}_{0}\frac{\sqrt{T/D}[\psi'(\frac{1}{2}+x)]^2dx/\sqrt{x}}{\Big[\ln(\frac{T_0}{T})-\psi(\frac{1}{2}+x)+\psi(\frac{1}{2})\Big]^2},
\end{equation}
where $x=Dq^2/4\pi T$ and we have introduced the characteristic skew-scattering time $\tau^{-1}_{\mathrm{sk}}=n_{\text{imp}}\alpha_{\text{so}}\nu^2_FV^3_0$. The remaining diagrams in the upper panel of Fig.~\ref{Fig-MT} can be evaluated in the same fashion \cite{Note1}. In the temperature range of interest we have $\ln(T_0/T)\gg1$, so that collecting all the terms, and extracting leading asymptotic expression we find for the total MT corrections in the skew-scattering mechanism: 
\begin{equation}\label{sigma-sk-MT}
\sigma^{\text{sk-MT}}_{xy}=-1.22\sigma_Q(\tau/\tau_{\mathrm{sk}})\sqrt{T/D}\ln^{-2}(T_0/T).
\end{equation}

There are six diagrams of the DOS type, each of which may have either two or four copies depending on the arrangements of the impurity lines, see the left panel of Fig.~\ref{Fig-DOS}. All these terms are structurally similar so we consider diagram-(a) as a guiding example and quote the results for all others. The response kernel is still given by Eq. \eqref{Q} but the self-energy part reads now as follows 
\begin{equation}
\Sigma^{\text{sk-DOS-a}}_{\bm{q}}=T\sum_{\varepsilon_n}\lambda^2_{\bm{q}}(\varepsilon_n,\Omega_{k-n})J^{\text{sk-DOS-a}}_{xy},
\end{equation}    
where 
\begin{align}
&J^{\text{sk-DOS-a}}_{xy}=n_{\text{imp}}\nu^3_F
\langle v_{\bm{p},x} v_{\bm{k},y}V_{\bm{p}\bm{k}}V_{-\bm{p}-\bm{k}'}V_{-\bm{k}'-\bm{k}}\rangle \nonumber \\
&\times\int d\xi_{\bm{k}'}G_{-\bm{k}'}(\varepsilon_{n+\nu})\nonumber \\
&\times\Big[\int d\xi_{\bm{k}} G_{-\bm{k}}(\varepsilon_{n+\nu})G_{\bm{k}}(\Omega_{k-n})G_{\bm{k}}(\Omega_{k-n-\nu})\Big]^2.
\end{align}
Repeating the same steps, we obtain for the corresponding conductivity correction \cite{Note2}
\begin{equation}\label{sigma-sk-DOS-a}
\sigma^{\text{sk-DOS-a}}_{xy} = -\frac{2e^2}{\sqrt{\pi}}\frac{\tau}{\tau_{\mathrm{sk}}}
\int\limits^{\infty}_{0}\frac{\sqrt{T/D}\psi'(\frac{1}{2}+x)\sqrt{x}dx}{\ln(\frac{T_0}{T})-\psi(\frac{1}{2}+x)+\psi(\frac{1}{2})}.
\end{equation}
An explicit calculation of the remaining diagrams in the left panel of Fig.~\ref{Fig-DOS} shows that the total DOS correction is equal to \cite{Note3}
\begin{equation}\label{sigma-sk-DOS}
\sigma^{\text{sk-DOS}}_{xy}=2.14\sigma_Q(\tau/\tau_{\mathrm{sk}})\sqrt{T/D}\ln^{-1}(T_0/T).
\end{equation}

In order to capture the skew scattering mechanism in the AL diagram one needs to go to the second loop order in interaction \cite{AL}. The corresponding diagram contains a quantum-crossing of a Hikami box that brings an extra smallness in $1/(p_Fl)\ll1$, as compared to both Eqs. \eqref{sigma-sk-MT} and \eqref{sigma-sk-DOS}. In addition, this AL term is more strongly suppressed in powers of $\ln(T_0/T)$, so that can be neglected. We should note, however, that in superconductors close to $T_c$ this contribution may be substantial due to its singular nature \cite{SL-AL}.  

\textit{Side-jump mechanism}.--We proceed with the analysis of the side-jump effect. This mechanism is manifested by an additional term in the matrix element of the velocity operator due to spin-orbit coupling
\begin{equation}\label{v-sj}
	\langle \bm{p}'|\hat{\bm{v}}| \bm{p}\rangle\!=\!\frac{\bm{p}}{m}\delta_{\bm{p}\bm{p}'}-\frac{i\alpha_{\text{so}}}{2mE_F}\!\sum_j 
	V_{\bm{p}-\bm{p}'}e^{i(\bm{p}-\bm{p}')\cdot\bm{R}_j}[\hat{z}\times(\bm{p}-\bm{p}')]
\end{equation}
where $\bm{R}_j$ is the radius vector of a given impurity. It generates additional diagrams and all the DOS processes are listed in the right panel of Fig.~\ref{Fig-DOS}. For the self-energy of the response kernel in Eq. \eqref{Q} one has, for diagrams (a) and (b),  
\begin{align}
\Sigma^{\text{sj-DOS-(a+b)}}_{\bm{q}}=T\sum_{\varepsilon_n}\lambda^2_{\bm{q}}(\varepsilon_n,\Omega_{k-n})J^{\text{sj-DOS-(a+b)}}_{xy},
\end{align}   
where 
\begin{align}
&J^{\text{sj-DOS-(a+b)}}_{xy}=-in_{\text{imp}}\nu^2_F\frac{\alpha_{\mathrm{so}}V^2_0}{2mE_F}\langle[\hat{z}\times(\bm{p}-\bm{p}')]_xv_{\bm{p},y}\rangle\nonumber \\ 
&\times\int d\xi_{\bm{p}} G^2_{\bm{p}}(\varepsilon_n)G_{\bm{p}}(\varepsilon_{n+\nu})G_{\bm{p}}(\Omega_{k-n}) \nonumber \\ 
&\times\int d\xi_{\bm{p}'} \Big[G_{\bm{p}'}(\varepsilon_{n+\nu})+ G_{\bm{p}'}(\varepsilon_n) \Big].
\end{align}
After all the technical steps we find an expression for the corresponding conductivity term which is structurally identical to Eq. \eqref{sigma-sk-DOS-a}, with the only difference in the numerical prefactor,
and also the ratio of scattering times should be replaced by the dimensionless side-jump parameter $\tau/\tau_{\mathrm{sk}}\to\varsigma_{\mathrm{sj}}=n_{\text{imp}}\alpha_{\mathrm{so}}\nu_FV^2_0/(\pi E_F)$.  Adding the remaining two terms from diagrams (c) and (d) we find in total  
\begin{equation}\label{sigma-sj-DOS}
\sigma^{\text{sj-DOS}}_{xy}=2.14\sigma_Q\varsigma_{\mathrm{sk}}\sqrt{T/D}\ln^{-1}(T_0/T).
\end{equation}
The side-jump scattering in MT diagrams, see Fig. \ref{Fig-MT}, does not give a finite contribution as respective terms (lower panel in Fig.~\ref{Fig-MT}) cancel each other, so that   
\begin{equation}
\sigma^{\text{sj-MT}}_{xy}=0.
\end{equation}
Likewise, we do not find finite contributions of side-jump processes in the AL diagrams, at least to the leading order in $1/(p_Fl)$.   

\textit{Summary and discussion}.--A few comments are in order in relation to results presented in this paper. We find that EEI in the Cooper channel produce temperature dependent corrections to the anomalous Hall conductivity. At low temperatures the effect is asymptotically dominated by the density of states processes in a parameter $\ln(T_0/T)$. Both skew scattering and side jump mechanisms contribute and scale mainly as a square root of temperature in three-dimensions with an additional logarithmic suppression. This may be relevant in regards to the recently reported measurements of AHE in HgCr$_2$Se$_4$ where $\sigma_{xy}\propto\sqrt{T}$ was observed \cite{Li-HgCrSe}. However, it must be carefully noted that effects of spin polarization may strongly influence the low temperature behavior of $\sigma_{xy}$. At temperatures below the Zeeman energy splitting $E_z$ of the conduction electron energies caused by the ferromagnetic polarization, we estimate that MT contribution in the skew-scattering mechanism is further suppressed in a parameter $(T/E_z)^{3/2}$, whereas DOS term is suppressed by $\sqrt{T/E_z}$. As a result, $\sigma^{\text{sk-MT}}_{xy}\propto T^2$ and $\sigma^{\text{sk-DOS}}_{xy}\propto T$, for $T< E_z$, while $\sigma_{xy}\propto \sqrt{T}/\ln(T)$ leading behavior pertains only for $T> E_z$.

In Fig. \ref{Fig-sigma} we attempted to fit the data with our analytical results. We find that both MT [Eq. \eqref{sigma-sk-MT}]  and DOS [Eq. \eqref{sigma-sk-DOS}] terms need to be retained for the best quantitative comparison. The relative importance of two extrinsic terms can be estimated from Eqs. \eqref{sigma-sk-DOS} and \eqref{sigma-sj-DOS} as $\sigma^{\text{sk}}_{xy}/\sigma^{\text{sj}}_{xy}\sim E_F\tau (\nu_FV_0)$.  For moderately strong impurity potential when, $\nu_FV_0\sim1$, skew scattering dominates in the metallic regime $E_F\tau\gg1$, however, both terms are of the same order close to Mott-Ioffe-Regel limit. The same analysis can be carried out for two-dimensional systems and we find much weaker temperature dependence $\sigma^{\text{sk}}_{xy}\simeq\sigma_Q(\tau/\tau_{\mathrm{sk}})\ln[\ln(T_0/T)/\ln(T_0\tau)]$. 

\textit{Acknowledgments}.--We are grateful to Yongqing Li and Kun Yang for discussions that stimulated this work. We thank Maxim Khodas for pointing Ref. \cite{Wolfle} to our attention. 
The experimental data presented in Fig. \ref{Fig-sigma} was made available to us by courtesy of Shuai Yang and Yongqing Li per Ref. \cite{Li-HgCrSe}. 
This work was supported by NSF Grants No. DMR-1653661 and No. DMR-1743986,  and the Ray MacDonald Endowment Award. 
This work was performed in part at Aspen Center for Physics, which is supported by National Science Foundation Grant PHY-1607611.

\end{document}